\documentclass[useAMS,usenatbib]{mn2e_mod}
\usepackage{natbib}
\usepackage{graphicx}
\usepackage{amsmath}
\usepackage{amssymb}
\usepackage{longtable}
\usepackage{url}
\usepackage{tabularx}

\usepackage{blindtext}
\newcommand  \kms      {\ifmmode {\rm km\,s}^{-1} \else km\,s$^{-1}$\fi}
\newcommand  \cc       {\hbox{cm$^{-3}$}}
\newcommand  \cmii     {\hbox{cm$^{-2}$}}

\newcommand  \ergs     {\ifmmode {\rm ergs\,s}^{-1} \else ergs s$^{-1}$\fi}
\newcommand  \ergcms   {\ifmmode {\rm ergs\,cm}^{-2}\,{\rm s}^{-1}
                        \else ergs\,cm$^{-2}$\,s$^{-1}$\fi}
\newcommand  \ergcmsA  {\ifmmode{\rm ergs\,cm}^{-2}\,{\rm s}^{-1}\,{\rm\AA}^{-1}
                        \else ergs\,cm$^{-2}$\,s$^{-1}$\,\AA$^{-1}$\fi}
\newcommand  \ergcmsHz {\ifmmode{\rm ergs\,cm}^{-2}\,{\rm s}^{-1}\,{\rm Hz}^{-1}
                        \else ergs\,cm$^{-2}$\,s$^{-1}$\,Hz$^{-1}$\fi}
\newcommand  \phcms    {\ifmmode {\rm ph\,cm}^{-2}\,{\rm s}^{-1}
                        \else ,ph\,cm$^{-2}$\,s$^{-1}$\fi}
\newcommand  \phcmsA   {\ifmmode {\rm ph\,cm}^{-2}\,{\rm s}^{-1}\,{\rm\AA}^{-1}
                        \else ph\,cm$^{-2}$\,s$^{-1}$\,\AA$^{-1}$\fi}
\def  \apj {ApJ}
\def  \apjl {ApJL}
 
\def  \mnras {MNRAS}

\def \aap   {A\&A}
\def \araa {AnnRevAstAp}

\def \nat  {Nature}

\def\micron{\ifmmode \mu{\rm m} \else $\mu$m\fi}
\def\mic{\ifmmode \mu{\rm m} \else $\mu$m\fi}
\def\kms{\ifmmode {\rm km\,s}^{-1} \else km\,s$^{-1}$\fi}
\def\Hubble{\ifmmode {\rm km\,s}^{-1}\,{\rm Mpc}^{-1}
        \else km\,s$^{-1}$\,Mpc$^{-1}$\fi}
\def\ergsec{\ifmmode {\rm ergs\;s}^{-1} \else ergs s$^{-1}$\fi}
\def\ergscm{\ifmmode {\rm ergs\,s}^{-1}\,{\rm cm}^{-2}
          \else ergs\,s$^{-1}$\,cm$^{-2}$\fi}
\def\ergscmA{\ifmmode {\rm ergs\,s}^{-1}\,{\rm cm}^{-2}\,{\rm \AA}^{-1}
          \else ergs\,s$^{-1}$\,cm$^{-2}$\,\AA$^{-1}$\fi}
\def\ergscmHz{\ifmmode {\rm ergs\,s}^{-1}\,{\rm cm}^{-2}\,{\rm Hz}^{-1}
          \else ergs\,s$^{-1}$\,cm$^{-2}$\,Hz$^{-1}$\fi}
\def\deg{\ifmmode ^{\rm o} \else $^{\rm o}$\fi}
\def\cc{\ifmmode {\rm cm}^{-3} \else cm$^{-3}$\fi}
\def\Msun{\ifmmode M_{\odot} \else $M_{\odot}$\fi}
\def\msun{\ifmmode M_{\odot} \else $M_{\odot}$\fi}
\def\msunyr{\ifmmode M_{\odot}\,yr^{-1} \else $M_{\odot}\,yr^{-1}$\fi}
\def\Msyr{\ifmmode M_{\odot}\,yr^{-1} \else $M_{\odot}\,yr^{-1}$\fi}
\def\Lsun{\ifmmode L_{\odot} \else $L_{\odot}$\fi}
\def\lsun{\ifmmode L_{\odot} \else $L_{\odot}$\fi}
\def\mseed{\ifmmode M_{seed} \else $M_{seed}$\fi}
\def\Mseed{\ifmmode M_{seed} \else $M_{seed}$\fi}
\def\qo{\ifmmode q_{0} \else $q_{0}$\fi}
\def\Ho{\ifmmode H_{0} \else $H_{0}$\fi}
\def\ho{\ifmmode h_{0} \else $h_{0}$\fi}
\def\qo{\ifmmode q_{0} \else $q_{0}$\fi}
\def\ao{\ifmmode a_{0} \else $a_{0}$\fi}
\def\to{\ifmmode t_{0} \else $t_{0}$\fi}
\def\Halpha{\ifmmode {\rm H}\alpha \else H$\alpha$\fi}
\def\ha{\ifmmode {\rm H}\alpha \else H$\alpha$\fi}
\def\Ha{\ifmmode {\rm H}\alpha \else H$\alpha$\fi}
\def\Pa{\ifmmode {\rm P}\alpha \else P$\alpha$\fi}
\def\Brg{\ifmmode {\rm Br}\gamma \else Br$\gamma$\fi}
\def\Hbeta{\ifmmode {\rm H}\beta \else H$\beta$\fi}
\def\hb{\ifmmode {\rm H}\beta \else H$\beta$\fi}
\def\Hb{\ifmmode {\rm H}\beta \else H$\beta$\fi}
\def\hg{\ifmmode {\rm H}\gamma \else H$\gamma$\fi}
\def\Hg{\ifmmode {\rm H}\gamma \else H$\gamma$\fi}
\def\Hgamma{\ifmmode {\rm H}\gamma \else H$\gamma$\fi}
\def\Hdelta{\ifmmode {\rm H}\delta \else H$\delta$\fi}
\def\Lya{\ifmmode {\rm Ly}\alpha \else Ly$\alpha$\fi}
\def\lya{\ifmmode {\rm Ly}\alpha \else Ly$\alpha$\fi}
\def\Lyb{\ifmmode {\rm Ly}\beta \else Ly$\beta$\fi}
\def\lyb{\ifmmode {\rm Ly}\beta \else Ly$\beta$\fi}
\def\hi{\ifmmode \mbox{{\rm H}\,{\sc i}} \else H\,{\sc i}\fi}
\def\hii{H{\sc ii}}

\def\heii{He{\sc ii}}

\def\ciii{\ifmmode {\rm C}{\sc iii}$]$ \else C{\sc iii}$]$\fi}
\def\civ{C{\sc iv}\,$\lambda1549$}

\def\ovi{O{\sc vi}}

\def\neviii{Ne{\sc viii}}

\def\mgii{Mg{\sc ii}}

\def\siiii{Si{\sc iii]}}

\def\feii{Fe{\sc ii}}

\def\o5007{[O\,{\sc iii}]\,$\lambda5007$}
\def  \Rin         {\hbox{$ {R_{\rm in}} $}}
\def  \Rout        {\hbox{$ {R_{\rm out}} $}}

\def  \Rhb         {\hbox{$ {R_{H \beta}} $}}
\def  \RME         {\hbox{$ {R_{\rm ME}} $}}

\def  \nh          {\hbox{$ {n_{\rm H}} $}}      
\def  \nH          {\hbox{$ {n_{\rm H}} $}}      
\def  \Ncol        {\hbox{$ {N_{\rm col}} $}}      
\def  \Cf          {\hbox{$ C_f $}}              
\def  \Cloudy      {{\it cloudy}}
\def  \cloudy      {{\it cloudy}}
\def \Lop      {${L_{5100}}$}
\def \Lopn      {${L_{5100,44}}$}

\def \LAGN{${L_{AGN}}$}
\def \Lion{${L_{ion}}$}

\def \Ledd{$L/L_{Edd}$}
\def \D4000{${\rm D_n4000}$}
\def \d4000{${\rm D_n4000}$}

\def  \d4000       {D_{n}4000}
\def  \kms         {\hbox{km s$^{-1}$}}          
\def  \ergs        {\hbox{erg s$^{-1}$}}              
\def  \cc          {\hbox{cm$^{-3}$}}

\def  \cmii        {\hbox{cm$^{-2}$}}
\def  \mic         {$\mu$m}

\def  \Prad        {{$P_{rad}$}}
\def  \Pgas        {{$P_{gas}$}}
\def  \Pmag        {{$P_{mag}$}}
\def  \La          {\ifmmode {\rm Ly}\alpha \else Ly$\alpha$\fi}
\def  \la          {\ifmmode {\rm Ly}\alpha \else Ly$\alpha$\fi}
\def  \Ka          {\ifmmode {\rm K}\alpha \else K$\alpha$\fi}
\def  \ka          {\ifmmode {\rm K}\alpha \else K$\alpha$\fi}
\def  \Lb          {\ifmmode {\rm L}\beta \else L$\beta$\fi}
\def  \Ha          {\ifmmode {\rm H}\alpha \else H$\alpha$\fi}
\def  \ha          {\ifmmode {\rm H}\alpha \else H$\alpha$\fi}
\def  \Hb          {\ifmmode {\rm H}\beta \else H$\beta$\fi}
\def  \hb          {\ifmmode {\rm H}\beta \else H$\beta$\fi}
\def  \Pa          {\ifmmode {\rm P}\alpha \else P$\alpha$\fi}
\def  \CIIIb       {\ifmmode {\rm C}\,{\sc iii]}\,\lambda1909
                     \else C\,{\sc iii]}\,$\lambda1909$\fi}
\def  \CIV         {\ifmmode {\rm C}\,{\sc iv}\,\lambda1549
                     \else C\,{\sc iv}\,$\lambda1549$\fi}
\def  \SIV         {\ifmmode {\rm Si}\,{\sc iv}\,\lambda1397
                     \else Si\,{\sc iv}\,$\lambda1397$\fi}
\def  \MgII         {\ifmmode {\rm Mg}\,{\sc ii}\,\lambda2798
                     \else Mg\,{\sc ii}\,$\lambda2798$\fi}
\def  \MGII         {\ifmmode {\rm Mg}\,{\sc ii}\,\lambda2798
                     \else Mg\,{\sc ii}\,$\lambda2798$\fi}
\def  \OVI         {\ifmmode {\rm O}\,{\sc vi}\,\lambda1035
                     \else O\,{\sc vi}\,$\lambda1035$\fi}

\def  \HeI        {\ifmmode {\rm He}\,{\sc i}\,\lambda5876
                     \else He\,{\sc i}\,$\lambda5876$\fi}
\def  \HeII        {\ifmmode {\rm He}\,{\sc ii}\,\lambda4686
                     \else He\,{\sc ii}\,$\lambda4686$\fi}
\def  \HeIIa        {\ifmmode {\rm He}\,{\sc ii}\,\lambda1640
                     \else He\,{\sc ii}\,$\lambda1640$\fi}
\def  \HeIIb        {\ifmmode {\rm He}\,{\sc ii}\,\lambda1085
                     \else He\,{\sc ii}\,$\lambda1085$\fi}
\def  \CII        {\ifmmode {\rm C}\,{\sc ii}\,\lambda1335
                     \else C\,{\sc ii}\,$\lambda1335$\fi}
\def  \CIII        {\ifmmode {\rm C}\,{\sc iii}\,\lambda977
                     \else C\,{\sc iii}\,$\lambda977$\fi}
\def  \CIIIb       {\ifmmode {\rm C}\,{\sc iii]}\,\lambda1909
                     \else C\,{\sc iii]}\,$\lambda1909$\fi}
\def  \CIV         {\ifmmode {\rm C}\,{\sc iv}\,\lambda1549
                     \else C\,{\sc iv}\,$\lambda1549$\fi}
\def  \bOIIIb       {\ifmmode {\rm [O}\,{\sc iii]}\,\lambda5007
                     \else [O\,{\sc iii]}\,$\lambda5007$\fi}
\def  \OIIIb       {\ifmmode {\rm O}\,{\sc iii]}\,\lambda1663
                     \else O\,{\sc iii]}\,$\lambda1663$\fi}
\def  \OVIII        {\ifmmode {\rm O}\,{\sc viii}\,653~{\rm eV}
                     \else O\,{\sc viii}\,$653~{\rm eV}$\fi}
\def  \OVII        {\ifmmode {\rm O}\,{\sc vii}\,568~eV
                     \else O\,{\sc vii}\,$568~{\rm eV}$\fi}
\def  \OVI         {\ifmmode {\rm O}\,{\sc vi}\,\lambda1035
                     \else O\,{\sc vi}\,$\lambda1035$\fi}
\def  \OIVb         {\ifmmode {\rm O}\,{\sc iv]}\,\lambda1402
                     \else O\,{\sc iv]}\,$\lambda1402$\fi}
\def  \OVb         {\ifmmode {\rm O}\,{\sc v]}\,\lambda1218
                     \else O\,{\sc v]}\,$\lambda1218$\fi}
\def  \bOIIb       {\ifmmode {\rm [O}\,{\sc ii]}\,\lambda3727
                     \else [O\,{\sc ii]}\,$\lambda3727$\fi}
\def  \bOIb       {\ifmmode {\rm [O}\,{\sc i]}\,\lambda6300
                     \else [O\,{\sc i]}\,$\lambda6300$\fi}
\def  \OI       {\ifmmode {\rm [O}\,{\sc i]}\,\lambda1304
                     \else [O\,{\sc i]}\,$\lambda1304$\fi}
\def  \NII         {\ifmmode {\rm N}\,{\sc ii}\,\lambda1084
                     \else N\,{\sc ii}\,$\lambda1084$\fi}
\def  \bNIIb         {\ifmmode {\rm [N}\,{\sc ii]}\,\lambda6584
                     \else [N\,{\sc ii]}\,$\lambda6584$\fi}
\def  \NIII         {\ifmmode {\rm N}\,{\sc iii}\,\lambda990
                     \else N\,{\sc iii}\,$\lambda990$\fi}
\def  \NIIIb         {\ifmmode {\rm N}\,{\sc iii]}\,\lambda1750
                     \else N\,{\sc iii]}\,$\lambda1750$\fi}
\def  \NIVb         {\ifmmode {\rm N}\,{\sc iv]}\,\lambda1486
                     \else N\,{\sc iv]}\,$\lambda1486$\fi}
\def  \NV          {\ifmmode {\rm N}\,{\sc v}\,\lambda1240
                     \else N\,{\sc v}\,$\lambda1240$\fi}
\def  \MgII        {\ifmmode {\rm Mg}\,{\sc ii}\,\lambda2798
                       \else Mg\,{\sc ii}\,$\lambda2798$\fi}
\def  \CVI        {\ifmmode {\rm C}\,{\sc vi}\,368~eV
                       \else C\,{\sc vi}\,368~eV\fi}
\def  \SiIV         {\ifmmode {\rm Si}\,{\sc iv}\,\lambda1397
                     \else Si\,{\sc iv}\,$\lambda1397$\fi}
\def  \bFeXb       {\ifmmode {\rm [Fe}\,{\sc x]}\,\lambda6734
                       \else [Fe\,{\sc x]}\,$\lambda6734$\fi}
\def  \MgX        {\ifmmode {\rm Mg}\,{\sc x}\,\lambda615
                       \else Mg\,{\sc x}\,$\lambda615$\fi}
\def  \MgXI        {\ifmmode {\rm Mg}\,{\sc xi}\,1.34~keV
                       \else Mg\,{\sc xi}\,1.34~keV\fi}
\def  \MgXII      {\ifmmode {\rm Mg}\,{\sc xii}\,1.47~keV
                     \else Mg\,{\sc xii}\,1.47~keV\fi}
\def  \bNeVb      {\ifmmode {\rm [Ne}\,{\sc v]}\,\lambda3426
                     \else [Ne\,{\sc v]}\,$\lambda3426$\fi}
\def  \bNeIIb      {\ifmmode {\rm [Ne}\,{\sc ii]}\,12.8\mu~m
                     \else [Ne\,{\sc ii]}\,$12.8 \mu m$\fi}
\def  \bOIVb      {\ifmmode {\rm [O}\,{\sc iv]}\,25.9\mu~m
                     \else [O\,{\sc iv]}\,$25.9 \mu m$\fi}
\def  \bNeVIRb      {\ifmmode {\rm [Ne}\,{\sc v]}\,14.3\mu~m
                     \else [Ne\,{\sc v]}\,$14.3 \mu m$\fi}
\def  \NeVIII      {\ifmmode {\rm Ne}\,{\sc viii}\,\lambda774
                     \else Ne\,{\sc viii}\,$\lambda774$\fi}
\def  \SiVIIa      {\ifmmode {\rm Si}\,{\sc vii}\,\lambda70
                     \else Si\,{\sc vii}\,$\lambda70$\fi}
\def  \NeVIIa      {\ifmmode {\rm Ne}\,{\sc vii}\,\lambda88
                     \else Ne\,{\sc vii}\,$\lambda88$\fi}
\def  \NeVIIIa      {\ifmmode {\rm Ne}\,{\sc viii}\,\lambda88
                     \else Ne\,{\sc viii}\,$\lambda88$\fi}
\def  \NeIX      {\ifmmode {\rm Ne}\,{\sc ix}\,915~eV
                     \else Ne\,{\sc ix}\,915~eV\fi}
\def  \NeX      {\ifmmode {\rm Ne}\,{\sc x}\,1.02~keV
                     \else Ne\,{\sc x}\,1.02~keV\fi}
\def  \SiXII        {\ifmmode {\rm Si}\,{\sc xii}\,\lambda506
                       \else Si\,{\sc xii}\,$\lambda506$\fi}
\def  \SiXIII      {\ifmmode {\rm Si}\,{\sc xiii}\,1.85~keV
                     \else Si\,{\sc xiii}\,1.85~keV\fi}
\def  \SiXIV      {\ifmmode {\rm Si}\,{\sc xiv}\,2.0~keV
                     \else Si\,{\sc xiv}\,2.0~keV\fi}
\def  \SXV      {\ifmmode {\rm S}\,{\sc xv}\,2.45~keV
                     \else S\,{\sc xv}\,2.45~keV\fi}
\def  \SXVI      {\ifmmode {\rm S}\,{\sc xvi}\,2.62~keV
                     \else S\,{\sc xvi}\,2.62~keV\fi}
\def  \ArXVII      {\ifmmode {\rm Ar}\,{\sc xvii}\,3.10~keV
                     \else Ar\,{\sc xvii}\,3.10~keV\fi}
\def  \ArXVIII      {\ifmmode {\rm Ar}\,{\sc xviii}\,3.30~keV
                     \else Ar\,{\sc xviii}\,3.30~keV\fi}
\def  \FeI_XVI      {\ifmmode {\rm Fe}\,{\sc 1-16}\,6.4~keV
                     \else Fe\,{\sc 1-16}\,6.4~keV\fi}
\def  \FeXVII_XXIII     {\ifmmode {\rm Fe}\,{\sc 17-23}\,6.5~keV
                     \else Fe\,{\sc 17-23}\,6.5~keV\fi}
\def  \FeXXV      {\ifmmode {\rm Fe}\,{\sc xxv}\,6.7~keV
                     \else Fe\,{\sc xxv}\,6.7~keV\fi}
\def  \FeXXVI      {\ifmmode {\rm Fe}\,{\sc xxvi}\,6.96~keV
                     \else Fe\,{\sc xxvi}\,6.96~keV\fi}
\def  \FeLa     {\ifmmode {\rm Fe}\,{\sc L}\,0.7-0.8~keV
                     \else Fe\,{\sc L}\,0.7-0.8~keV\fi}
\def  \FeLb     {\ifmmode {\rm Fe}\,{\sc L}\,1.03-1.15~keV
                     \else Fe\,{\sc L}\,1.03-1.15~keV\fi}
\def\hi{\ifmmode \mbox{{\rm H}\,{\sc i}} \else H\,{\sc i}\fi}
\def\hii{H\,{\sc ii}}

\def\heii{He\,{\sc ii}}

\def\ciii{\ifmmode {\rm C}\,{\sc iii]} \else C\,{\sc iii]}\fi}
\def\civ{\ifmmode {\rm C}\,{\sc iv} \else C\,{\sc iv}\fi}
\def\cv{\ifmmode {\rm C}\,{\sc v} \else C\,{\sc v}\fi}

\def\ovi{O\,{\sc vi}}

\def\mgii{Mg\,{\sc ii}}

\def\feii{Fe\,{\sc ii}}

\def  \L46         {$ L_{46} $}

\def  \Ledd        {$ L/L_{\rm Edd} $}    
\def  \MBH         {$ M_{\rm BH} $}     
\def  \M*         {$ M_{*} $}     
\def  \m9          {$ m_9 $}
\def  \nh          {$ n_{\rm H} $}      
\def  \N10         {$ N_{10} $}
\def  \Ncol        {$ N_{\rm col} $}    

\title[Testing BLR models with reverberation mapping]
{Testing broad line region models with Reverberation mapping }

\author[Hagai Netzer]
{Hagai Netzer $^1$\thanks{E-mail: hagainetzer@gmail.com} \\
$^1$School of Physics and Astronomy, Tel-Aviv University, Tel-Aviv 69978, Israel
}

\begin{document}
\maketitle

\begin{abstract}
New reverberation mapping (RM) measurements, combined with accurate luminosities and line ratios, provide
strong constraints on the location of the line emitting gas in the broad line region (BLR) of active galactic nuclei (AGN). 
In this paper I present new calculations of radiation pressure and magnetic pressure confined clouds and apply them to a ``generic AGN'' and to 
 NGC\,5548. 
The new calculations are in good agreement with the observed lags of all broad emission lines, and with the luminosities of \La, \civ\,1549, 
\ovi\,1035 and \heii\,1640.
They are also in reasonable agreement with the luminosities of \mgii\,2798 and the 1990\AA\ blend of \ciii\ and \siiii\ lines for high metallicity gas. 
They explain the changes in time-lag following an increase in continuum luminosity and their dependencies on the inner and outer boundaries of the BLR. They also 
predict very strong Balmer and Paschen continua with  
important implications to continuum RM experiments.
However, the calculated Balmer and Paschen line luminosities are too weak, by factors of 2-5. This ``Balmer line crisis'' was  noted in several earlier works and  is now
confirmed and constrained by RM measurements that were not available in the past.
It seems that present photoionization codes that use the escape probability formalism, 
 fail to correctly compute the Balmer line luminosities in high density, large optical depth gas.

\end{abstract}

\begin{keywords}

(galaxies:) quasars: general; (galaxies:) quasars: supermassive black holes; galaxies: nuclei; galaxies: actives
\end{keywords}

\section{Introduction}

High density gas on a sub-pc scale, next to active black holes (BHs) in the centers of active galactic nuclei (AGN), has been studied for years (see e.g. detailed review and numerous
 references in \citealt{Netzer2013}). 
The time variable emission lines from the broad line region (BLR), together with the variable
optical-UV-X-ray continuum, 
are the most recognized signatures of activity in type-I AGN, those sources with a clear view of the vicinity of the BH.
Observations of strong, broad, semi-forbidden lines, such as \ciii\,1909,  
suggest that the lowest density of the gas is about \nh=10$^{9-10}$ \cc, where \nh\ is the hydrogen particle density. 
The upper limit on the density is not well determined and densities as high as 10$^{13}$ \cc\ have been proposed. .

Progress in reverberation mapping (RM) of the BLR gas provides an opportunity to map the location and motion of the gas, and to estimate the BH mass (\MBH). 
RM measurements are now available for a large number of broad emission lines:  
\ha, \hb, \La, \civ1549, \heii\,1640, \heii\,4686, \mgii\,2798, and several blends of \feii\ lines 
\citep{Kaspi2000,Bentz2010,Bentz2013,Hu2015,Du2015,Grier2017,Pei2017,Lira2018,Kriss2019}. They have been used to derive the ``mean emissivity radius'' (\RME) 
and the ``mean responsivity radius'' of many lines, and to demonstrate that these radii increase roughly in 
proportion to \Lop$^{1/2}$, where \Lop\ is the monochromatic continuum luminosity at 5100\AA\  in \ergs. 
For sources in the range $10^{44}<$\LAGN$<10^{46}$\,\ergs, where \LAGN\ is the bolometric luminosity of the source, the mean emissivity
radius of the \hb\ line (hereafter \Rhb) is about 34$[$\Lopn$]^{1/2}$ light days (ld), where \Lopn=\Lop/$10^{44}$\,\ergs\
 \citep{Kaspi2000,Bentz2013}. RM studies also show different \RME\ for  different lines, starting from the 
\heii\ lines (about  20-50\% of \Rhb), 
going through  \la\ and \civ1549\  (about 50-80\% of \Rhb), other Balmer lines (similar but not always identical to \Rhb), and to the \feii\ and \mgii\ lines 
(somewhat larger than \Rhb\ with a large uncertainty). 

Dust RM experiments, comparing V and K-band luminosity variations, as well as K-band interferometry, provide additional information about the location of the outer boundary of the BLR, \Rout,
identified with graphite grains sublimation radius and the inner walls of a central dusty torus.
Earlier theoretical ideas suggested a well defined sublimation radius associated with the inner walls of the central dusty torus,
\cite[e.g.][and references therein]{Netzer2015}. However, it is now clear that grain size and composition play important roles in setting \Rout\
\cite[see the detailed study by][]{Baskin2018}. Moreover, the exact torus geometry plays an important role too and
makes it difficult to convert the delayed K-band emission to \Rout\ \cite[e.g.][]{Stalevski2012,Goad2015}.
However, the good agreement between the dust location derived from  RM measurements \cite[][]{Koshida2014}, and the location inferred from the K-band interferometry by GRAVITY (Dexter et al. 2020), 
indicate that geometry and complicated light-echo effects may not be very important. The observations obtained
so far suggest \Rout/\Rhb$\approx 4$. 
This  \Rout/\Rhb\  is somewhat larger than the one calculated by \cite{Baskin2018} for the case of isotropic emission (\Rout/\Rhb$=1.5-3$). 
Since \Rout\ depends on \LAGN\ and \Rhb\ depends on \Lop, there is an additional ambiguity due to uncertainties on the  
spectral energy distribution (SED) of such sources, especially the spectral shape of the ionizing continuum.

In this paper I present new, global BLR photoionization models and confront them with measured luminosities and mean emissivity radii of several broad emission lines.
Most calculations pertain to an ``RM-sample AGN'': an object representing those sources with reliable, multi-season RM measurements. 
Of the few available global BLR models, none takes into account, in detail, the role of radiation pressure force
and none attempts to explain both the gas distribution, as derived from RM measurements, and the luminosities of the strongest lines.  
As argued below, radiation pressure confined (RPC) cloud models provide the best agreement between line luminosity and line lag except for the Balmer lines. This may indicate a   
fundamental difficulty in the calculations of these lines.  
In \S2 I describe current BLR cloud models. In \S3 I present new photoionization calculations 
and in \S4 I compare them with AGN observations. \S5 presents a discussion of the new findings with emphasis on the ``Balmer lines crisis''. 
Throughout this work I assume a standard cosmological model with $\Omega_{\Lambda}=0.7$, $\Omega_{m}=0.3$, and
$H_{0}=70\,\kms\,\, {\rm Mpc}^{-1}$.

\section{Global BLR models}

Three generic cloud models have been proposed, over the years, to describe the space and density distribution of the gas in the BLR: magnetic pressure confined (MPC) clouds,
locally optimally emitting clouds (LOC), and RPC clouds.
Most of this work concerns with the case of {\it pure} cloud models, where only one type of clouds occupy the entire volume.
Several continuous flow wind-type models have also been proposed but without specific calculations of the emitted spectrum. Such models will not be discussed here.
 
Given \LAGN, continuum SED, and gas metallicity,  the complete definition of a cloud model requires 6-8 parameters.
Five of these  
 are common to all models: the inner and outer radii, \Rin\ and \Rout, the column density of the clouds, \Ncol, and 
two parameters that define the radial dependence of the global covering factor, \Cf,
\begin{equation}
    d C_f=c_1 R^{-p} dR \, ,
    \label{eqn1}
\end{equation}
where $c_1$ is the covering factor normalization.
The line luminosity is obtained by computing the line efficiency coefficient per unit covering
factor, $\epsilon(line,R)$, such that $dL(line,R) \propto \epsilon(line,R) d C_f$.

In reality, \Rout\ is the dust sublimation radius which is determined by the source luminosity, grain size, and metallicity
\cite[see][]{Netzer1993,Netzer2015,Baskin2018}, and is thus not a free parameter. 
In this work I assume \Rout=4\Rhb\ (see  discussion below).
 I also fix the inner BLR boundary to \Rin=0.2\Rhb\ which is consistent with the shortest observed emission-line time-lag. 
The exact value affects mostly the luminosity of lines from
highly ionized species, like \ovi\,1035, and the change in \RME\ in response to continuum variations (\S4 below).   
As detailed below, between two and three additional parameters are required to characterize the gas density, \nH, and/or gas pressure, \Pgas.

The purpose of this work is to  calculate the mean emissivity radius (\RME), and luminosity, of various emission lines and compare them with the results of 
RM experiments and BLR spectroscopy. This is easily done if 
 $\epsilon (R)$ is a powerlaw in radius, $\epsilon \propto R^{q}$.
Here $q=0$ represents the case
where the emitted flux in the line is proportional to the incident continuum flux 
in exactly the same way at all distances\footnote{
$\epsilon(R)$ can be viewed as a linear responsivity factor analog to
the powerlaw responsivity $\eta (R)$ used in      
other papers ($L_{line} \propto L_{cont.}^{\eta(R)}$). $q=0$ corresponds to $\eta = 1$}.
 Given this definition, and defining $t=p-q$,
\begin{equation}
{\rm R_{ME} } =  \frac {  \int_{R_{in}}^{R_{out}} R^{1-t} dr } {  \int_{R_{in}}^{R_{out}} R^{-t} dr } = 
  R_{in} \frac {1-t}{2-t}
   \frac {  
        [(R_{out}/R_{in})^{2-t}   -1]   } {  [(R_{out}/R_{in})^{1-t}   -1]          
    }      \, , 
\label{eqn2}
\end{equation}
for  $t \neq 1$ and $t \neq 2$.
As shown in several earlier works \cite[e.g.][]{Goad2015}, and in the following sections of this paper,
 such a power-law dependence of the line efficiency 
on $R$ is highly simplified and there is no replacement for real, step-by-step calculations of the
emitted spectrum. The mean emissivity radius of the {\it total} line and continuum emission
is, however, easier to calculate since in this case $q=0$ and $t=p$. 
As shown below, information  from the measured luminosities and lags of \La, \civ\,1549, the Balmer lines and the Balmer continuum, all  suggest that in many AGN $p \geq 2$.
Note also that \RME\ depends both on \Rin\ and \Rout, an important issue which is discussed below.

MPC clouds have been suggested by \citet{Rees1989} and further discussed in  
\citet{Netzer1990,Goad1993,Kaspi1999,Bottorff2002,Netzer2015,Lawther2018}.
Such clouds are confined by the pressure of the external magnetic field, \Pmag, 
 assumed to originate  from the central accretion disk (AD). If \Pmag$\gg$\Prad, we get  \Pgas=\Pmag.
The properties of magnetically confined clouds are very similar to the hot-gas confined clouds discussed in \cite{Stern2016}.
All published BLR models investigated so far assumed a simple radial dependence of the pressure,
 \Pmag$= c_2 R^{-s}$, where $c_2$ is the (unknown) pressure normalization and 
$s$ is in the range 0-2. Thus,  
the total number of parameters in the MPC cloud model, not counting \Rout, is 6. 

MPC clouds are confined on all sides and there is no specific limit on their column density. Some MPC  models are normalized in such a way that 
the column density is large enough  to make the back of the cloud almost completely neutral \cite[e.g.][]{Netzer1990,Kaspi1999}. 
Others (Goad et al. 1993; \citealt{Lawther2018}) allow 
matter bounded clouds that are optically thin to the Lyman continuum radiation. The additional assumption that the clouds retain their mass as they move in or out, 
provides a way to define \Ncol\ as a function of radius.
 For gravitationally bounded MPC clouds, $v_{cloud} \propto R^{-1/2}$ and $p=2/3s-3/2$ (Rees et al. 1989) reducing the number of free parameters to 5.
 Detailed applications of the MPC cloud model to the 
specific case of NGC\,5548, are discussed in \cite{Kaspi1999}, and  \cite{Lawther2018}.

LOC models have been proposed by \cite{Baldwin1995} and studied in numerous publications \cite[e.g.][]{Korista1997,Korista2000,Bottorff2002,Goad2015,Korista2019}. 
The model assumes a range of densities at any given location and is based on the idea that the escaping radiation from a given location is dominated by 
line and continuum emission from clouds whose density is close to the density of the highest line production efficiency. 
The column density of the clouds in most of the published models is constant, in the range $10^{23}-10^{24}$\,\cmii, and
the clouds are not confined. The local density in the LOC model is assumed to be distributed over
a large range, $ \sim 10^{7}-10^{14}$~\cc, and is defined by a power-law distribution with 
three parameters: \nh(min), \nh(max) and the  power-law index which in most LOC models is set to -1. 
Given this prescription, the number of parameters in LOC models, not counting \Rout, is 7.

RPC models have been proposed by \cite{Dopita2002} and  have been applied first to the dusty, narrow line region (NLR) gas. 
In this case, the luminosity of the central source, assumed to be the only source of external pressure,
 provides radiation pressure force that compresses the dusty gas to a density 
where \Prad=\Pgas\
at $\tau$(Lyman)$\sim 1$.
The resulting density is proportional to $R^{-2}$ which gives a constant ionization parameter: $U=Q$(Lyman)/($4 \pi R^2 c)$, 
where $Q$(Lyman) is the total number of ionizing
photons per second and $c$ is the speed of light. 
For the dusty NLR gas, $ \log U \sim -1.5$. Radiation pressure compression is not important in those locations where \Pgas$(R)>$\Prad$(R)$.

\cite{Baskin2014} (hereafter BLS14) applied the radiation pressure confinement idea to large column density, dust-free BLR clouds.
They showed that in all locations where \Prad$\gg$\Pgas\ at the illuminated face of the cloud, 
L/Ledd$<1$, and the column  density is large enough to keep the material at the back of the cloud neutral,
the gas motion is dominated by gravity  and the clouds maintain a simple hydrostatic structure. 
 Since \Prad$\propto R^{-2}$, this is the case for {\it all clouds}, regardless of their distance from the central source, provided
 the ambient density is low enough. In this model, the ionization parameter inside the clouds,  
 near the hydrogen ionization front, is $\log U \sim -1$. Further study of this model is given in \cite{Stern2016}.

 The structure of RPC clouds change significantly when the column density drops below a certain value which, for the SED 
 considered by BLS14, is $few \times 10^{23}$. The radial structure and the motion of such clouds is determined by the combines
 effects of radiation pressure and gravity. Examples of such  motion are given in \cite{Netzer2010}.
 For a large enough \Ncol, the number of parameters in the RPC model not counting \Rout\ is 5.
 
BLS14 did not address the case where high density gas in some parts of the BLR result in \Pgas$>$\Prad. 
They also did not consider the case where the local properties of the clouds are 
determined by some combination of \Prad\ and \Pmag, or the case where there is not enough gas, in some directions, to result in optically thick RPC clouds.
 They  made no attempt to match RPC cloud
models to RM measurements but line luminosities were roughly compared with typical equivalent widths in order to estimate \Cf.

The main premise of the RPC model contradicts some of the assumptions of both the LOC and MPC models. 
In the LOC case, gas whose density is such that \Prad$>$\Pgas,
and its column density large enough, will be compressed on a short time scale and its contribution to the local line emission will differ,
substantially,  
from what is assumed by the model. 
This is also the case in the MPC clouds model in those locations where \Prad(R)$\geq$\Pmag(R).

The recent progress in RM experiments (\S1) resulted in a large number of AGN with known BH mass, luminosity and lag for several emission 
lines, and graphite sublimation radius.
Such observations can be used to put
strong observational constraints on the distribution of the BLR gas and to test the various global cloud models. 
Several attempts of this type, focusing on the well studied case of NGC\,5548 and using the MPC  and LOC cloud models, have 
been published \cite[][]{Kaspi1999,Lawther2018,Bottorff2002,Korista2019}. However, none included the inevitable influence of the
radiation pressure force.
Below I use new photoionization calculations to explore various RPC and MPC cloud models.
I do not include LOC models which are inconsistent
with the basic assumption of the RPC model.

\section{Multi-cloud Photoionization Models}

\subsection{Spectral energy distribution and gas composition}

The present calculations apply to a generic RM-sample AGN whose broad emission line spectrum is the mean of the population in terms
of line luminosity and line time-lag. This includes  sources with \Lopn=0.1-10.   
The central source of radiation is assumed to be an optically thick accretion flow, either a geometrically thin AD like the ones calculated by \cite{Slone2012}
 but without a disk wind, or a yet unexplained flow with a different SED. The mass of the central BH is assumed to be \MBH=$10^8$\,\msun\ but this plays a minor
role in the present work.

Four different SEDs are considered. Two represent standard geometrically thin ADs with different accretion rates and spin parameter ($a$): one with   
$a=0.7$ and radiation conversion efficiency  $\eta=0.104$ (hereafter AD1)
 and one with $a=-1$ and  $\eta=0.038$ (hereafter AD2). The accretion rates are determined by the requirement that 
\Lop=$10^{44}$\,\ergs. This results in \Rhb=34\Lopn$^{1/2}$\,ld. The corresponding Eddington ratios (\Ledd) are 0.108
and 0.045.  

The two disk SEDs are combined with a 0.5-100 keV X-ray powerlaw continuum with an energy index of $\alpha_X$=0.9. 
The optical-to-X-ray index ($\alpha_{OX}$) is 1.38 for AD1 and 1.55 for AD2.
None of the results pertaining to the RM-sample AGN are sensitive to the exact X-ray properties. 
Given these SEDs, the mean energies of the ionizing photons are 2.16 Rydberg for AD1 
and 1.8 Rydberg for AD2. The ionizing luminosities are $10^{45}$\,\ergs\  and $10^{44.34}$\,\ergs, respectively.

For proper comparison with earlier BLR models and SEDs, I also considered two broken powerlaw SEDs with properties similar to those
considered in the literature. The first is characterized by a $L_{\nu} \propto \nu^{-1.6}$ Lyman continuum and an IR-optical-UV continuum
which is a combination of powerlaws with slopes
0.5 and 1. The X-ray continuum is similar to the one assumed for the accretion disks. The fourth SED is the one observed for NGC\,5548
during 2014 
 \citep{Mehdipour2015} with 
\Lion=$10^{44.34}$ \ergs, Q(H)=1.29$\times 10^{54}$, \LAGN=$10^{44.44}$\,\ergs, and \Lop=$10^{43.37}$\,\ergs.

\begin{figure}
\includegraphics[width=9cm]{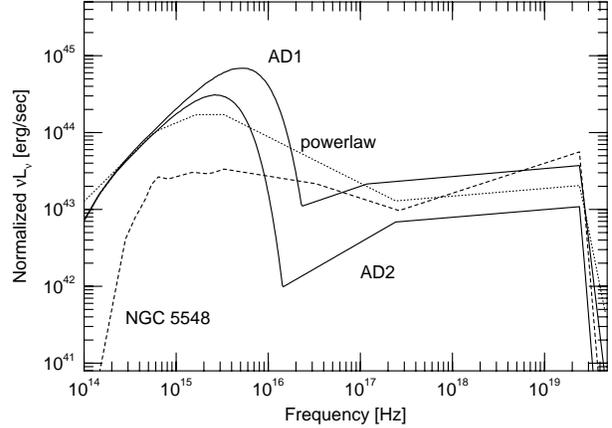}
\caption{
Various SEDs used in this work. 
}
\label{fig1}
\end{figure}

Table~\ref{tab:table1} provides information about the different SEDs and 
Fig.~\ref{fig1} shows all four of them. The table lists \Lion/\Lop, mean energies of ionizing photons in Rydberg, and $\alpha_{OX}$.
The first of those is very important since most of the properties discussed in this work are determined by \Lion\ yet the common normalization in
RM studies is relative to \Lop.

\begin{table}
\centering
\caption{SED properties}
\begin{tabular}{lccc}
SED    & \Lion/Lop\ & Mean E(Ryd) & $\alpha_{OX}$ \\
\hline
AD1      & 10        & 2.16  &  1.38  \\
AD2      & 2.2       & 1.80  & 1.55   \\
Powerlaw & 3.4       & 3.20  & 1.39  \\
NGC\,5548& 9.2         & 7.8   & 1.27 \\
\end{tabular}
\label{tab:table1}
\end{table}

Line luminosities  depend also  on the gas composition.
In the present work this is taken to be one of the default AGN compositions provided by \Cloudy:\\
$(H:He:C:N:O:Ne:Mg:Al:Si:S:Ar:Ca:Fe)=$\\
$10^{-4}\times (10^4:10^3:
 2.45:0.85:4.9:1.0:0.347:0.0234:0.347:0.177:0.025:0.023:0.282)  $\\
One of the cases presented below assumes metallicity which is three times higher. For all elements, excluding nitrogen and helium, this is obtained by simply
multiplying the abundance by 3. For nitrogen which is a secondary element, I multiply the above number by 6.4.

\subsection{Photoionization codes}

Most of the models presented below were calculated using version C17.01 of the code \Cloudy\ \cite[][]{Ferland2017}. BLS14 used version 10 of the code. 
They assumed a low starting density, 
well below the density where \Prad=\Pgas, 
and integrated until the material becomes completely neutral. This is reached at a column density of $few \times 10^{23}$\,\cmii. 
In the present calculations I used \Ncol=10$^{23.5}$. The exact value of \Ncol\ makes 
little difference to the line intensities provided the back of the cloud is more than 95\% neutral. While the column density is very large, 
the ionized parts of the clouds are significantly smaller, with \Ncol=$10^{23}$\,\cmii\ or less. 
Thus, the 
Compton depth, and the optical depth at the various hydrogen bound-free edges, do not depend much on the exact choice of \Ncol.

A problem which was encountered by BLS14, is the disruption of the clouds by the internal line radiation pressure. This instability, which
is also encountered in \Cloudy\ C17.01, is not very important since RPC 
clouds are not confined in the lateral direction and hence will likely change their shape, on a sound speed crossing time, to compensate for this effect. 
However, the instability can cause  
the calculations to crush. Because of this, BLS14 shut-off the internal line ration pressure (by using the command {\it no radiation pressure}). Experimenting 
with RPC models that do not result in such crush shows that the internal density and temperature structures of the clouds are considerably different, 
but the emergent line flux is hardly affected. All this is discussed and demonstrated in BLS14. 
Most cases shown below are taken from \Cloudy\ calculation without including line radiation pressure.

Since \Cloudy\ is basically the only code used in recent years to calculate BLR models, I used the code ION, most recently described in \cite{Mor2012}, 
to complement and verify the results.  
The current version, ION2019, contains all the physical processes included in \Cloudy\ except for the treatment of molecular gas. The atomic data set is less complete but is 
fully updated for the lines considered here and for the main coolants of the gas. The hydrogen atom scheme used in ION2019 contains fewer levels but 
the number is large enough, and the conditions far enough from LTE, such that the emergent hydrogen spectrum is very similar to the one calculated by \Cloudy. 
A more fundamental issue is the treatment of HeI and HeI-like 
ions where the \Cloudy\ scheme contains many more levels and a full {\it n-l} treatment. This is not included in ION2019 and hence the resulting HeI spectrum is 
less reliable. These lines are not included in the present work and are also not very important in controlling the gas temperature. ION includes also the treatment of line radiation 
pressure but the process does not cause any instability, probably because the number of lines, and the exact treatment of photon escape, are different from those used 
in \Cloudy. Here, again, 
I verified that shutting off this process makes only a small difference to the emergent line flux.

\begin{figure}
\includegraphics[width=9cm]{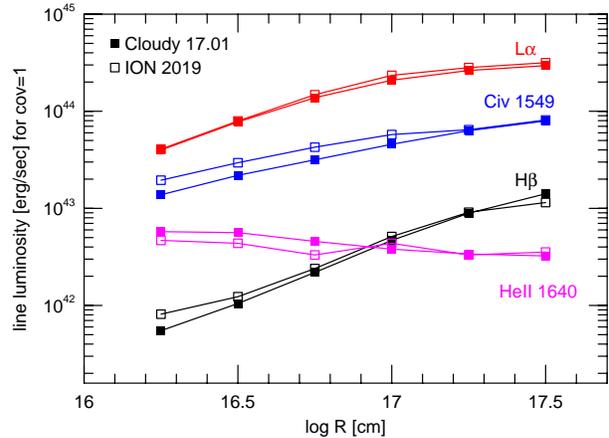}
\caption{
Comparison of \Cloudy\ C17.01 and ION2019 calculations for the  standard RPC AD1 cloud model
considered in this work. 
}
\label{fig2}
\end{figure}

Fig.~\ref{fig2} shows a comparison of the \hb, \La, \civ\,1549 and \heii\,1640 calculations made by the two codes for the basic RPC-AD1 model described below,
over a large range of distances. 
The differences in all cases are below 0.15 dex and there is no systematic trend. A comparison of several MPC cloud models, not shown here, show them
to be in very good agreement too.
The only significant difference ($\sim 0.2$ dex) is found for the \civ\,1549 line in the AD2 model. 
 I have also compared the density and temperature run inside the cloud and found very small differences. The small difference in the Lyman and Balmer lines 
between the two codes are 
probably due to the somewhat different expressions used to calculated the local escape probability for the lines. These are well know issues which are not directly  related 
to the more fundamental 
uncertainties associated with the hydrogen line transfer discussed in \S5 below. 
For the rest of this paper, all results shown are those calculated
using \Cloudy.

BLS14 did not address the comparison of the integrated line emission, and mean emissivity radii, with observations. They only show   
line intensities at various distances from the central AD for \Cf=0.3. 
According to BLS14, the observed \la\ and \civ1549\ luminosities are consistent with this covering factor 
but this is not the case for \hb\ whose calculated equivalent width (EW) is a factor 
3-4 below the mean observed value. This issue is not discussed  further in their work except for commenting on 
the difficulty in calculating reliable Balmer line intensities. 

The difficulty in 
reproducing the observed luminosity of \hb\ and other Balmer lines has been addressed by \cite{Kaspi1999} in their MPC modeling of the spectrum of NGC\,5548.
 This was attributed
to inadequate treatment of such lines in photoionization models where radiative transfer is based on the local escape probability method.
 \cite{Lawther2018} repeated most of the
\cite{Kaspi1999} calculations and applied them to the  2014 data
of NGC\,5548. They also added a detailed discussion of the bound-free diffuse continuum (hereafter DC) which was not included in \cite{Kaspi1999},
and noted the  that the calculated L(\hb) is much weaker than observed.
As shown below, the new RPC and MPC calculations show a similar discrepancy
with \hb\ and \ha\ observations.

\subsection{New RPC and MPC calculations}

The aim of the present work is to use observed line luminosities, and line lags, in order to test and constrain RPC and MPC cloud models. 
The emphasis is on {\it pure} models, i.e. those where either \Prad$\gg$\Pmag\ or \Pmag$\gg$\Prad. More realistic BLRs would probably contain clouds of both types, for example RPC clouds 
in the inner BLR and MPC clouds in the outer BLR.

As explained, RPC cloud models require 5 parameters for a given \LAGN, SED and gas composition. All the radial parameters can be expressed relative to \Rhb, defined as 
the RM distance of the \hb\ line: \Rhb=34$[$\Lopn$]^{1/2}$ ld.
The total line luminosities are obtained by integrating 
from \Rin=0.2\Rhb\ to \Rout=4\Rhb\ and  the  line emission is assumed to be strongly suppressed, by the torus dust, at larger distances. 
For each case I used eqn.~\ref{eqn1} with a large range of the covering factor parameter 
$p$, between 1 and 3.5, normalized to give a total covering factor of 0.4. This was considered to be the 
largest covering factor which is still consistent with the assumption of no shadowing of one cloud by another.  

For each model  and each line, I calculated
the mean emissivity radius which is assumed to represent the measured
RM distance of the line. 
As explained in various earlier works \cite[e.g.][]{Bottorff2002}, the mean emissivity radius 
can differ from the responsivity distance which is more closely related to the peak of the cross correlation function (CCF) between the line and continuum light curves.
The mean emissivity radius depends on line responsivity and 
also on the duration of the driving continuum pulse \cite[e.g.][]{Goad2015}. Here I assume that the mean measured RM distances in the RM-sample are
based on long duration driving continuum events and are therefore similar to the mean emissivity weighted radii.

There are several differences between the new MPC calculations presented here and the earlier ones applied to NGC\,5548. 
The first is the constant pressure assumption (some earlier calculations assumed constant gas density). The second is the lower limit
on the gas density at the illuminated face of the clouds
imposed by the radiation pressure force. To take this into account, I assumed  
\Pmag$ \propto R^{-s}$, and adjusted the density at the illuminated face such that  \Pmag$>$\Prad\ at \Rin. 
I also assumed $s<2$ thus \Pmag$>$\Prad\ throughout  the 
BLR. This normalization is not the only possibility and there can be cases where \Prad\ dominates over the inner
BLR and \Pmag\ further out.  Such cases were not considered. 
An additional difference from the earlier calculations is the neglect of the requirement that individual clouds retain their mass as they move in or out. This requirement sets the 
column density as a function of distance and also allowed some clouds to become optically thin close to the central source \cite[for details see][]{Netzer1990}.
All clouds in the present 
paper have the same large column density of $10^{23.5}$\,\cmii. The covering factor dependence is set solely by the parameter $p$ in eqn.~\ref{eqn1}

All MPC cloud models presented below  assumes $s=1$.  The relatively 
small change in gas temperature at the illuminated face over a large range of distances shows that, to a good approximation, \nh$\propto R^{-0.95}$ and
 $U \propto R^{-1.05}$.

\section{Comparison with Observations}

The results presented here pertain to two cases:  RM-sample AGN and the specific case of NGC\,5548.

\subsection{RM-sample AGN}

This set of calculations is aimed to compare RM results, and broad emission line luminosities, with the mean properties of the RM-sample AGN presented
in \cite{Kaspi2000} and \cite{Bentz2013}.
The mean observed line luminosities and lags are collected from the literature and are listed in Table~2.
The main references are:
\cite{Netzer1990}, \cite{Netzer1995}, \cite{Kaspi2000}, \cite{Telfer2002},
\cite{Kim2010}, \cite{Bentz2010}, \cite{Bentz2013}, \cite{Lira2018}, \cite{Pei2017} 
 and \cite{Kriss2019}.
The line luminosities are normalized to L(\hb), and the line lags to \Rhb/$c$.
They show a  large scatter which is mostly intrinsic and related to different phases of activity, different SED shapes, different continuum variability time scales,
disk inclinations and accretion rates. The table lists also the mean luminosity of the \Pa\ line and the following calculations add
the predicted \Brg\ line. This is necessary in order to compare with coming GRAVITY observations that can measure both of these lines in the
K-band. The listed time-lag for the line is not measured by RM experiments but 
rather deduced from
the similarity of the \hb\ and \Pa\ line profiles.   

In general, the line intensity ratios do not depend strongly on \LAGN\ (no line reddening is assumed); 
an important issue which was discussed in numerous earlier publications \cite[e.g.][and references therein]{Netzer1979,Netzer1995,Baron2016} and 
addressed in \S5 below. 

\begin{table}
\centering
\caption{
Mean observed broad emission line luminosities and distances for RM-sample AGN with  monochromatic luminosity of \Lop=$10^{43}-10^{45}$\,\ergs.
}
\scalebox{0.8}{%
\begin{tabular}{lcccc}
Line                 & Luminosity           & Relative   & distance       & Relative   \\
                     & (\ergs)               & intensity  & (ld)           & distance                 \\      
\hline
\hb\                 &  0.018 \Lop\         &  1      &34$[$\Lopn$]^{1/2}$&    1                \\
\la\                 &                      & 5-20       &                &   0.4-0.8           \\ 
\ha\                 &                      & 2-4        &                &  1-1.3           \\
\Pa\                 &                      & 0.3-0.4    &                &  1(?)            \\
\civ1549\            &                      &4-15        &                &   0.4-0.8           \\
\heii\,1640\         &                      &0.5-2       &                &   0.2-0.5            \\
\mgii\,2798\         &                      &0.5-3       &                &   1-1.5             \\
\ciii,\siiii\,1900   &                      &2-4       &                &     ?                \\
\ovi\,1035           &                      &1-3         &                &     ?              \\
Graphite dust        &                      & 100-400    &                & 3 - 4               \\
\hline
\end{tabular}}
\label{tab:table2}
\end{table}

The new calculations pertain to two different scenarios: fixed boundaries BLRs and changing boundaries BLRs.

\subsubsection{Fixed boundaries BLRs}
This set of calculations assumes that \Rin\ and \Rout\ do not change as the source luminosity increases or decreases.
For the RPC clouds I follow the approach of BLS14 and start the integration into the cloud assuming a density well below the critical density 
corresponding to \Prad=\Pgas. This
translates to $U=1-10$. 
The calculations stop at \Ncol=$10^{23.5}$\,\cmii\ but this number is somewhat arbitrary since the ionized column in all cases does not exceed $10^{23.3}$\,\cmii.
The covering factor of individual clouds are very small (the {\it open geometry} option in \Cloudy) and no 
obscuration of one cloud by another is considered.
The covering factor parameter, $p$, is allowed to vary in the range 1-3.5. The larger this parameter is the smaller is the emissivity weighted radius. 
In general, $p$ is the most important parameter
in determining the mean emissivity radii of the lines. 

\begin{figure}
\includegraphics[width=9cm]{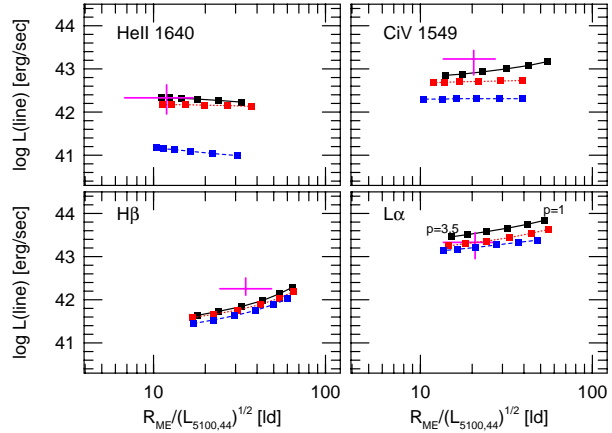}
\caption{Luminosities and mean emissivity radii for four BLR lines, \Cf=0.4 and a range of covering factor parameter $p$
(marked along the lines and decreasing from $p=3.5$ (left) to $p=1$ (right)).
{\it Solid lines:} RPC model, AD1 SED. {Dashed lines:} RPC model , AD2 SED. {\it Dotted lines:} MPC model, AD1 SED.
The estimated observational uncertainties are given by the central cross in each panel.
Note that none of the models can reproduce the mean observed L(\hb) despite the large covering factor.
The weak ionizing continuum AD2 SED fails also to explain L(\heii\,1640) and L(\civ\,1549).
} 
\label{fig3}
\end{figure}

Fig.~\ref{fig3} shows results  for the three main models: RPC-AD1, RPC-AD2 and MPC-AD1.
In all cases \Cf=0.4.    
The calculations  are compared with the luminosities and lags shown in Table~\ref{tab:table2} (marked as magenta crosses in all panels). 
Table~\ref{tab:table3} provides more information for those values of $p$ which results in the closest agreement with \Rhb: 2.4 for the RPC-AD1 model and 2.3 for the MPC-AD1 model. 
It also shows calculations for $p=1.2$ that results in longer lags for all lines and and for a RPC model with metallicity which is three times solar.
Given the observational 
uncertainties, and intrinsic scatter, all the results in the range $p=2.2\pm 0.5$ are consistent with all mean lags.
The time lags for the Balmer and Paschen continua and \ovi\,1035 are not known. However,  their luminosities per unit covering factor are
almost constant with distance, similar to the case of \heii\,1640 (Fig.~\ref{fig2}). This means that the line efficiency factor, $\epsilon(R)$, depends very weakly on $R$. Because of this, the calculated lags are also similar to those of the \heii\ lines.

I have also calculated RPC models for the powerlaw SED shown in Fig.~\ref{fig1}. The results are very similar to
those of the AD1 SED in all respects. The preferred
covering factor parameter in this case is somewhat smaller,  $p=2.2$, which reflects the fact that the ionizing luminosity of this SED, 
 which dominates the radial distribution of the clouds,  is somewhat smaller than in AD1 (see Table~\ref{tab:table1}). 
The results of this case are not discussed further in this work. 

Unlike the time lags, there is a serious disagreement, which is clearly visible in Fig.~\ref{fig3} and Table~\ref{tab:table3},
between the observed and calculated \hb\ luminosity, 
regardless of the value of $p$. A similar discrepancy is found for \ha, \Pa\ and \hg\ (not shown in the diagram).
 BLS14 studied the case of \Cf=0.3 and did not calculate mean emissivity radii. They also noted the \hb\ deficiency. 
A similar problem was found in the earlier NGC\,5548-MPC calculations of \cite{Kaspi1999} and also in \cite{Lawther2018}.
Below I refer to this discrepancy as the ``Balmer lines crisis''.

The covering factor used in this work is the largest which is still consistent with the assumption of no cloud obscuration. 
The models assume isotropic ionizing 
continuum yet some BLRs are likely to form
a rotating system around the mid-plane of the AD where the ionizing photon flux can be very different
(see e.g. the new GRAVITY observations of 3C\,273 by \cite{Sturm2018}). This could increase the
 discrepancy between the observed and the calculated
Balmer line luminosities. The exact magnitude of this effect depends on BH mass and spin \cite[see][figures 7 \& 8]{Laor1989}.
 An additional factor which is not addressed here is the uneven emission from the illuminated and dark sides of the clouds.

A major finding of this work is that, contrary to the Balmer lines crisis, 
there is no luminosity crisis for \La, \civ\,1549, the two \heii\ lines (note that \heii\,1640/\heii\,4686$\approx 10$) and
\ovi\,1035. In fact, the calculated 
luminosities are consistent with a large range of covering factor from 0.1 to 0.4. 
The calculated luminosities of the \mgii\,2798 blend, and the 1900\AA\ blend (a combination of \ciii\,1909 and \siiii\,1893), in solar composition gas,
are considerably smaller than the ones observed. Much of this discrepancy is cured when higher metallicity is used. Unfortunately,
there are no reliable measurements of the lags of these lines.

\subsubsection{Changing boundaries BLRs}
A large increase in \LAGN\ can result in dust sublimation and a different outer boundary 
provided the sublimation time can be neglected \citep{Baskin2018}.
Such an event results in additional line emission and a change in the mean emissivity radius. 
The effect on \Rin\ is more difficult
to assess, in particular in those cases where the time scale of the variations is shorter than the dynamical time at \Rin.
One possibility is that radiation bounded clouds close to \Rin\ will become partly transparent thus reducing the efficiency
of the lines from the more neutral species. 

\begin{figure}
\includegraphics[width=9cm]{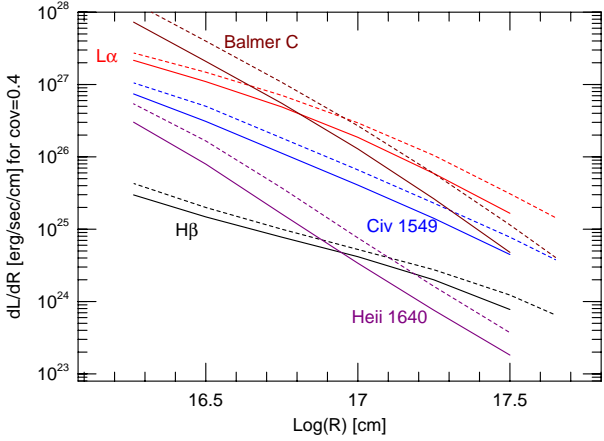}
\caption{Line and bound-free continuum luminosities in response to an AD1 SED with $p=2.4$: \Lop=$10^{44}$\,\ergs\ (solid lines)
and \Lop=$2 \times 10^{44}$\,\ergs\ (dashed lines). 
} 
\label{fig4}
\end{figure}

 In this section
I investigate  RPC and MPC cloud models where \LAGN\ increases by a factor 2, \Rout\ increases by $2^{1/2}$, 
and there is no change in \Rin. Since the two types of models lead to similar results, I focus on RPC models.
Fig.~\ref{fig4} shows $dL/dR$ for several lines, and the Balmer continuum, in the $p=2.4$ RPC case,
 before (solid lines) and after (dashed lines) such an increase.   
 The lines emission after the change start at the same \Rin\ and extend
 all the way to the new \Rout. The integrated old and new luminosities and lags are listed in Table~\ref{tab:table4}.

\onecolumn
\begin{table}
\centering
\caption{
Calculated luminosities and lags relative to L(\hb)=0.018\Lop\,\ergs\ and \Rhb=34$[$\Lopn$]^{1/2}$ ld, for 
RPC and MPC models with AD1 SED. Numbers 
in parentheses are for 3$\times$solar metallicity gas. 
}
\scalebox{0.8}{%
\begin{tabular}{lccccccccccc}

\hline
                      &\hspace{0.5cm}  RPC $p=2.4$&    &&\hspace{0.5cm}RPC $p=1.2$&  &&\hspace{0.5cm} MPC $p=2.3$&  &&\hspace{0.5cm} MPC $p=1.2$&              \\ 
\hline
\hline
Line                  & Relative   & Relative & &Relative  &Relative &&Relative  &Relative&&Relative  & Relative   \\
                      & luminosity & distance & &luminosity&distance &&luminosity&distance&&luminosity& distance      \\
\hline
\hb\                  & 0.38 (0.36)      &  1   (1)       & &0.91      &1.79     &&  0.36    &  1     &&0.75      &1.77     \\
\la\                  & 20.8 (19.7)      & 0.75 (0.76)    & &34.6      &1.43     &&  13.6    &  0.80  &&21.9      &1.48      \\
\ha\                  & 0.97 (0.88)      & 1.16 (1.16)    & &2.72      &1.93     &&  0.78    & 1.12   &&1.81      &1.86     \\
\Pa\                  & 0.06 (0.056)     & 1.16 (1.14)    & & 0.17     &1.97     &&  0.07    & 1.14   &&0.17      &1.90     \\
\Brg\                 & 0.009 (0.009)    & 0.82  (0.82)   & & 0.016    &1.73     &&  0.015   & 0.83   &&0.025     &1.59        \\
\civ1549\             & 5.5  (4.63)      & 0.70  (0.61)   & &8.5       &1.43     &&  2.82    &  0.54  &&2.95      &1.02     \\
\heii\,1640\          & 1.2  (0.93)      & 0.44  (0.45)   & &1.0       &0.87     &&  0.81    &  0.49  &&0.77      &0.96     \\
\mgii\,2798\          & 0.21 (0.35)      & 1.06  (0.99)   & &0.52      &1.9      &&  0.52    &  1.20  &&1.32      &1.97     \\
\ciii+\siiii\,1900    & 0.45 (0.99)      & 1.49  (1.24)   & &1.64      &2.1      &&  0.18    & 1.18   &&0.45      &1.95     \\
\ovi\,1035            &2.61  (2.06)      & 0.49  (0.48)   & &2.58      &1.0      &&  0.06    & 0.29   &&0.03      &0.38      \\
Balmer cont.          &32.0  (30.9)      & 0.48  (0.48)   & &30.6      &0.92     &&  39.8    &  0.53  &&40.4      &0.99     \\
Paschen cont.         &22.4  (20.2)      & 0.45  (0.45)   & &20.1      &0.87     &&  24.6    &  0.48  &&22.4      &0.88     \\
\hline
\end{tabular}
}
\label{tab:table3}
\end{table}

\begin{table}
\centering
\caption{Luminosities and lags due to a luminosity increase by a factor 2.
In both cases the SED is AD1 and $p=2.4$.
Line luminosities and lags are relative to the nominal values for \hb\ listed in Table~\ref{tab:table3}.
}
\begin{tabular}{lcccc}

\hline
                   &\Lop=$10^{44}$\,\ergs\ && \Lop=$2 \times 10^{44}$\,\ergs\           \\
\hline     
\hline
Line               &Luminosity  &Lag        & Luminosity   & Lag      \\
\hline
\hb\              & 0.38        &  1        & 0.55      & 1.24       \\
\la\              & 20.8        & 0.75      & 30.9     &  0.93    \\
\civ1549\         & 5.5         & 0.70      & 8.9     &   0.81      \\
\heii\,1640\      & 1.2         & 0.44      & 2.35      & 0.47  \\
Balmer cont.      & 32.0        & 0.48      & 62     &    0.51   \\
\hline
\end{tabular} 
\label{tab:table4}
\end{table}

\begin{table}
\centering
\caption{
Observed and calculated luminosities and lags for the RPC model of NGC\,5548. Luminosities are relative to L(\hb)=$4.5 \times 10^{41}$\,\ergs\
and lags relative to 16 ld.  
The assumed dust sublimation radii are: \Rout(1)=49 ld and  \Rout(2)=86 ld.  
}
\begin{tabular}{lccccccc}
\hline
                  &           &\Rout(1) $p=1.8$  &           & & \Rout(2)  $p=1.8$  &         \\
\hline     
\hline
Line              &Observed  &Calculated  &Calculated& & Calculated   & Calculated   \\
                  &luminosity&luminosity  &lag       & & luminosity   & lag      \\
\hline
\hb\              &  1      & 0.18        &  1       & &  0.21        &  1.4      \\
\la\              & 8.96    & 5.5         & 0.77     & &  5.7         &  0.98    \\
\ha\              & 3.7     & 0.55        & 1.27     & & 0.72         & 1.96      \\
\civ1549\         & 8.96    & 4.2         & 0.67     & &  4.3         &  0.84     \\
\heii\,1640\      & 1.10    & 1.1         & 0.44     & &  1.1         &  0.52     \\
\mgii\,2798\      & 1.5     & 0.21        & 1.07     & &  0.25        &  1.59      \\
Balmer cont.      &         & 11.5        & 0.31     & &  11.3        &  0.36      \\
Paschen cont.     &         & 8.7         & 0.37     & &  8.6         &  0.42     \\
\hline
\end{tabular}
\label{tab:table5}
\end{table}


\twocolumn

There are two important things to note. The local (same $R$) response of different lines is very different representing the different emission coefficient $\epsilon(R)$
(\S2). Most noticeable
is the large difference between the small increase in  L(\hb) compared
with the much larger changes in L(\heii\,1640) and L(Balmer continuum). 
Unfortunately, it is not at all clear whether the change in L(\hb) can be trusted given the problem in calculating, reliably,  the intensity of this line.

A second and related aspect is the very different change in \RME\ for the different lines. The largest relative changes are
in those lines whose luminosity per unit covering factor increase outwards (\hb\ and \La, see also Fig.~\ref{fig2}). Even for these lines, the increase in \RME\
is smaller than the naively expected change by a factor $2^{1/2}$. This is related to the specific value of $p=2.4$ which emphasizes the inner parts of the BLR, and to
the fact that \Rin\ was not allowed to change (eqn. 2).  Thus L(\heii\,1640) and L(Balmer continuum) increase by a factor of $\sim 2$ at all distances but their mean emissivity radii
hardly change.

Models with smaller $p$, e.g. the case of $p=1.2$ shown in Table~\ref{tab:table3}, give more weights to the
outer
parts of the BLR. In such cases, the changes in \RME\ are larger and approach $2^{1/2}$. However, such models fail
to reproduce the mean emissivity radii of \La, \civ\,1549, \heii\,1640\ and the Balmer lines. 
All these predictions could be tested against observations.

\subsection{The BLR in NGC\,5548}

The most detailed studies, so far, involve the intermediate luminosity AGN NGC\,5548. This object was the target of
three very large optical-UV campaigns, in 1989, 1993 and 2014, and  a large number of additional space-borne
and ground-based
 campaigns \cite[see][and references therein]{Pei2017,Korista2019,Kriss2019}.
Except for several years of very low luminosity, which will not be discussed here, the   SED  
is dominated by a strong X-ray continuum  which is substantially  
different from the three other SEDs studied in this work. Details and a plot of this SED are provided in \S3.1. 

The large number of RM campaigns provide an opportunity to compare 
the RPC cloud model for this source with published LOC 
and MPC models.
Extensive discussions of the
BLR response in this source, in the framework of the LOC model, are given in  \cite{Goad2015} and  \cite{Korista2019}.
These studies addressed density, column density and luminosity dependencies as well as the effect of the dusty torus location 
on the observed line luminosities and lags.  
They also addressed the luminosity and lag of the bound-free continuum.
For various MPC models see \cite{Kaspi1999} and \cite{Lawther2018}.
  
The RPC clouds model presented here were applied to the luminosity and SED of NGC\,5548 during 
2014. 
I used the line and continuum observations described in \cite{Lawther2018}, \cite{Pei2017} and \cite{Kriss2019} and assumed the SED properties detailed in \S3.1.  
Using the \cite{Bentz2013} normalization, \Rhb$\approx 16$ ld, about twice  the lag measured during the STORM 2014 campaign and close to the 25-year mean for this source
 \cite[][]{Pei2017}. 
According to \cite{Korista2019}, the much shorter lags (a factor of $\sim 2$) observed in 2014 are consistent with the very different continuum variability time scale during this campaign 
(see Fig.~7 in their paper).

 Using the above SED, I calculated
 emissivity-weighted radii and luminosities for the strong BLR lines, and the bound-free continua, under two
 assumptions  about the dust sublimation radius: \Rout=49 ld and \Rout=86 ld. These bracket the estimated \Rout\ based on dust RM. 
These values are 2-3 times smaller than the radius considered by
\cite{Goad2015} and \cite{Korista2019} who investigated different BLR and torus geometries, and the changes of the sublimation radius in response to continuum variations.
The LOC models used in these papers assumed $p=1.2$. The main reason for the very large \Rout, and the small $p$, is the too small L(\hb) predicted by the model compared
with the observations. These assumptions resulted in stronger Balmer lines and smaller L(\La)/L(\hb).
As before, \Cf=0.4.

The results of the new calculations are presented in Table~\ref{tab:table5} alongside the mean observed line luminosities
taken from the 1989 and 2014 optical-UV campaigns. I chose $p=1.8$ which gives \Rhb=16 ld for
 \Rout=49; the mean observed lag for this \Lop. For \Rout=86 ld and $p=1.8$ I find \Rhb=22 ld.  The calculated luminosities and lags are in good agreement
 with the observations
for \La, \civ\,1549 and \heii\,1640. However, this is not the case for the Balmer lines and the \mgii\,2798 blend. 
The lags for these lines are in good agreement with the observations but the luminosities are about a factor 4-5 
too small. 
This is in line with the Balmer lines crisis mentioned earlier.

I have also experimented with super-solar metallicity, increasing the metal abundance by a factor 3. This
increased the \mgii\,2798 luminosity by about a factor 2 but made little difference for the other lines and continua.

\section{Discussion}

\subsection{A comparison of RPC and MPC models}
 
The new calculations presented here pertain to simplified scenarios where {\it all clouds} in the BLR are controlled either
my radiation pressure or by external magnetic pressure. Real BLRs probably include some mixture of the two, depending on distance from the
BH, position relative to the central disk, etc. 
 I also neglected the possibility that some clouds, especially
at small distances, are not ionization bounded.
Nevertheless, identifying the main features of the two {\it pure} cases is important for the understanding of the BLR physics.

Perhaps the most important feature of the RPC clouds model is  the fact that compression by radiation pressure force
is unavoidable when \Prad$\geq$\Pgas. This sets the run of temperature and density across the cloud and hence the emergent line and continuum flux. 
It also results in a smaller number of parameters compared with the LOC and MPC cloud models. 
Given radiation pressure force, most earlier LOC and MPC models must be modified
in a significant way. In particular, in typical AGN SEDs, the radiation pressure force restricts the range of allowed 
densities in pure LOC and MPC models. For example, in the RM-sample model with the AD1 SED, the lowest allowed density for which
\Pgas$>$\Prad\ is about $10^{11}[(7 \,\, ld)/R]^2$\,\cc. Gas with lower density will be compressed  on a short time scale 
and, perhaps, blown out of the system. 

Large column density RPC clouds are characterized by an almost constant $U$ at the hydrogen ionization front, regardless of $R$. 
This behaviour can explain the great spectral similarity between AGN of very different luminosities, and the different lags of 
some of the lines like
the Balmer lines, \La, \civ\,1549 and the \heii\ lines. An additional feature is the large range of ionization parameters inside the clouds.
This allows to obtain strong lines of highly ionized and less ionized species from the same location. For example, it  helps to explain the good agreement
between the predicted and observed luminosity of the \ovi\,1035 line. Unfortunately, the mean emissivity radius of this line
has never been observed by RM experiments and hence cannot be used to test this prediction.

The situation regarding the  luminosity of low ionization lines like 
 \mgii\,2798 and the optical \feii\ lines, and their (rather uncertain) lags, is not as clear. The observed line luminosities
seem to indicate  a drop of $U$ with distance from the central source, which is less  consistent with the RPC model. 
This is not entirely clear because of the dependence on gas metallicity. It may also indicate that realistic BLR models must include a mixture
of RPC and MPC clouds.

RPC cloud models require a narrow range of covering factor parameters, $p \approx 2.2 \pm 0.5$. Smaller values that were used in the past 
seem to be in contradiction
with line and dust RM measurements and with K-band interferometry. 
This range of $p$ can explain the measured lags of {\it all} broad lines with reliable RM measurements.

Magnetic pressure confined clouds have been investigated in several earlier works (Rees et al.; 1989, Goad et al. 1993;
\citealt{Kaspi1999}, \citealt{Lawther2018}). 
Here the number
of free parameters  is 5 or 6, depending on whether or not the
 clouds retain their identity as they
move around the BH.  The present calculations are  different from the earlier ones in two ways. First, $p$ is larger than in earlier works. Second, 
I only consider MPC clouds  for which 
\Pmag$\gg$\Prad\ at the illuminated face. 
For the case demonstrated here, with \Pmag$\propto R^{-1}$, the line luminosities and lags are similar to those of the RPC 
clouds model for some of the lines (\heii\,1640, \La, \civ\,1549) and different for others (\mgii\,2798). The largest difference is for the \ovi\,1035 line blend 
which is under-luminous in the present MPC models because of the lower ionization parameter. A similar effect was noted by \cite{Stern2016} who calculated the luminosity
of the \neviii\,774 blend in cases where \Prad\ is the dominated external pressure and compared them with cases where external, hot gas pressure is more important. 
Obviously, there is a range of possible radial dependencies of \Pmag\  that are different from the one considered here.
There is also an uncertainty associated with the exact \Prad/\Pmag\ at different locations
inside the BLR.

In conclusion, the main advantage of RPC cloud models is the smaller number of model parameters and the fact that radiation pressure must operate at
all locations, even if not as the dominant source of pressure. 
Both
RPC and MPC models can reproduce the observed lags for all broad emission lines with reliable RM measurements
and neither can explain the observed luminosities of the Balmer lines.

\subsection{What is the origin of the Balmer lines crisis?}

For $p=2.2\pm0.5$, 
both the calculated luminosities and lags of \La, \hii\,1640\ and \civ\,1549 are consistent with 
the observations. The agreement is not as good for the L(\mgii\,2798) but most of the discrepancy can be cured by resorting to
higher metallicity gas.  In contrast,  
the calculated \ha, \hb\ and \Pa\ luminosities are too week by factors of 2-5 despite the very good agreement between the observed and the calculated mean emissivity 
radii.
A similar discrepancy was found by \cite{Kaspi1999} in their study of NGC\,5548, and later confirmed by 
 \cite{Lawther2018}. The new calculations verify and extend this finding to  more physically consistent models.

BLR clouds are very optically thick in many Balmer and Paschen lines. In the models considered here, $\tau$(\hb)=$few \times 10^3$. Unlike resonance lines,
 like \La, where most of the scattering of the line photons is taking place close to their place of origin, the non-resonance 
Balmer line photons can propagate though
the cloud before escaping the gas. This can affect the $n>1$ level populations, and the local gas temperature, in a way that is difficult to simulate 
properly by models like \Cloudy\ and ION which use the local escape probability formalism.

It is worth noting in passing that the relative intensities of \ha, \hb\ and \Pa\ are similar to what is predicted from LTE calculations with 
a Boltzmann level populations.
This was noted in earlier works which discussed the observed Balmer line ratios in AGN in the context of the Boltzmann plot often used to analyze optically thin
plasma  \citep{Popovic2003,Ilic2012}. The inclusion of \La\ clearly shows a large discrepancy between such calculation and the observed hydrogen line spectrum;  
the direct result of the
very large optical depth of the lines. 

A related issue is the luminosities of the various bound-free continua. In the models calculated here, the optical depth at the Balmer limit is less than 10 and 
the one at the Paschen limit less than 1. Thus transfer is not likely to be important. Some confirmation of this is the measured L(Balmer cont.)/L(\La) which
was found by \cite{Wills1985} to be of order unity, similar to the RPC and MPC model predictions listed in Table~\ref{tab:table3}.

In conclusion, it  seems that the escape of Balmer and Paschen
line photons from high density, large optical depth BLR clouds, is more efficient than calculated by \Cloudy\ and by ION. Other lines and continua 
are less affected.

The discrepancy discussed here is related to the long-standing problem of the relative intensity of the Lyman and Balmer lines in AGN spectra
 \cite[see e.g.][and references therein]{Netzer1979,Netzer1995}.
The conclusions reached here are based on more physically motivated photoionization models and  supported by additional observational  constraints provided by 
numerous RM studies not available for the earlier studies.

Alternative explanations for the too strong (compared with calculations) \hb\ line, or equivalently too small \la/\hb,
involve line and/or continuum reddening \cite[e.g.][]{Netzer1995,Baron2016}. This
cannot solve the entire problem because reddening involves more luminous
intrinsic ionizing continuum which results in larger \RME\ for many lines.

It is hard to estimate what will be the effect of including a more realistic transfer method on the intensities of other emission lines. 
In particular, how will the energy redistribute inside the cloud. 
The lines that are most likely to deviate from what was found here are \mgii\  and \feii\ lines that are produced in the low ionization parts of the clouds.
 Lines that originate from the
more highly ionized parts are in much better agreement with the observations and, perhaps, less likely to be affected. 

Finally, a comment on the mean emissivity radius of the {\it total} emission from the BLR. This can be estimated, quite accurately, by noting that all
the ionizing radiation is absorbed by the large column density clouds considered here, as well as some of the incident radiation between 0.25 and 1 Ryd.
This is almost independent of the cloud locations. The calculations show that for
 $p=2.4$, \RME(total BLR emission)$\approx 0.5$\RME(\hb). This is a factor 6-8 smaller than \Rout. Thus, a prediction of the RPC cloud model
is that most of the flux emitted by the clouds is at very small distances compared
 with the dust sublimation radius. 
A large fraction of this radiation is due to various bound-free continua, which are discussed in the next section.

\subsection{Diffuse continuum emission from the BLR}

Recent studies clearly indicate that long wavelength continuum variations lag
the variable UV continuum in a wavelength dependent way.
 This was interpreted in several publications as a delayed response of a centrally illuminated
accretion disk to luminosity variations of the central source. Modeling suggests that the derived AD size is some 2-4 times larger than
the size of standard ADs such as the ones assumed here 
\cite[e.g.][and references therein]{Edelson2019}. Such ideas have been challenged by new observations, and new calculations, 
of the diffuse continuum (DC) from the BLR. Several recent BLR calculations show that  such emission 
can cause much, perhaps all of the observed 
effect \citep{Lawther2018,Chelouche2019,Korista2019}.
Thus the origin of the DC is crucial for the understanding of both the BLR physics and the properties of the central disk.

An important new result is that \RME(Balmer continuum) and \RME(Paschen continuum) in the RPC and MPC cloud models are significantly
smaller than \Rhb. For $p=2-2.5$ they correspond to $\sim0.5$\Rhb.
For the covering factor considered here (0.4) and the AD1 and AD2 continua,
the DC contribution to the total emitted 2500-10000\AA\ flux  
is very significant, ranging between 30 and 90\%.
 This is similar to what has been proposed by Chelouche et al. 2019 for the case of MK\,279. 
The relative DC contributions for the powerlaw and NGC\,5548 SEDs are smaller because in these cases
the optical continuum is stronger relative to ionizing continuum which determines
  the DC luminosity. This explains
much of the difference between the relative DC contributions calculated here and the ones
 shown in \cite{Lawther2018} and \cite{Korista2019}.

\begin{figure}
\includegraphics[width=9cm]{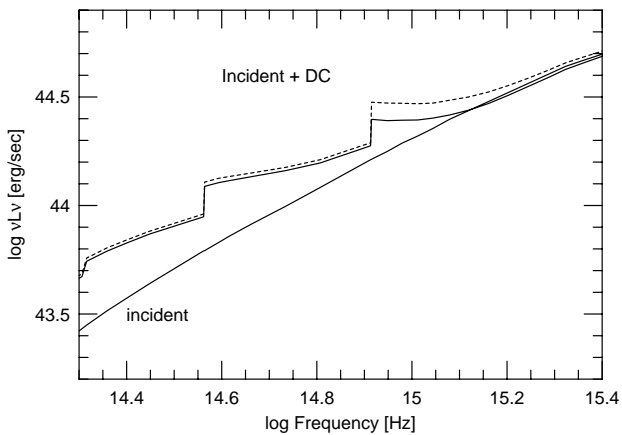}
 \caption{Diffuse continuum emission from a RPC-dominated BLR. In this case
 the SED is AD1, \Lop=$10^{44}$\,\ergs, the cloud is at \RME(Balmer continuum) and the covering
 factor is 0.4. The bottom solid line is the incident AD continuum, the middle solid line is the observed emission from a spherical cloud distribution,
and the upper dashed line is a side view of the total emission for a flat rotating BLR (broad emission lines are not plotted for clarity).
 }
 \label{fig5}
\end{figure}

Fig.~\ref{fig5} shows the DC continuum for the generic RM-sample AGN.
In this case I assumed that a single RPC cloud at a distance of 17 ld (\RME(Balmer continuum) in Table~\ref{tab:table3}) from 
an AD with \Lop=$10^{44}$\,\ergs, is  an adequate representation
of the entire DC emission. 
The diagram shows the incident AD1 continuum and two possibilities for the DC spectrum,
one representing a spherical cloud distribution (middle solid line) and one a thick disk-like BLR 
seen from the side (upper
dashed line). The differences are due to the fact that in the spherical situation, both the incident continuum and the diffuse emission are attenuated on
their way to the observer.
 As seen from the diagram, the DC adds about 40\% to the AD continuum at 3000\AA, about 60\% at 5000\AA\ and about 90\% at 7700\AA. 
 Naively, such a DC flux would result in a combined AD+DC continuum time lag at 5000\AA\ of
about 6$[$\Lopn$]^{1/2}$ days 
if the AD variations are wavelength independent (see more explanation in Chelouche et al. 2019). The combined continuum time lags are significant
even at smaller covering factors. 
Given these assumptions, 
 the lags are similar to the optical continuum lags reported in \cite{Edelson2019}, \citep{Fausnaugh2017},
and other cases. 

While the role of DC emission from the BLR must be studied more carefully in individual sources, and calibrated against observed line 
and continuum intensities and SED shapes, 
it is important to emphasize that such DC contributions, and lags, are unavoidable. 
Moreover, in low luminosity AGN, neglecting to account for this emission will lead to an unrealistically strong stellar contribution to the total observed continuum.

\section{Conclusions}
This paper presents new photoionization  models in attempt to reproduce the observed luminosities and lags of the strongest broad emission lines in AGN.
There are 4 different SEDs and two types of models: 
 radiation pressure confined (RPC) clouds and magnetic pressure confined (MPC) clouds.
The main results are:
\begin{enumerate}
\item
The observed lags of \La, \hb, \ha, \civ\,1549, and \heii\,1640 are reproduced by both RPC and MPC cloud models
provided the covering factor parameter $p$ ($dC_f \propto R^{-p} dR$) is in the range 2-2.5. This is
also the case for the \mgii\,2798 and \feii\ lines whose lags are more uncertain.
\item
The observed luminosities of \La, \civ\,1549, \heii\,1640, and \ovi\,1035, are well reproduced by RPC clouds models with ADs and powerlaws SEDs. The MPC clouds model
is not as successful in reproducing the luminosities of the higher ionization lines. 
The luminosities of \mgii\,2798 and the 1900\AA\ blend of \ciii\ and \siiii\ lines
approach their observed values for super solar metallicity gas. 
\item
The calculated Balmer and Paschen line luminosities fall short of the observed luminosities by factors of 2-5. This is interpreted as a failure of 
the escape probability formalism used in \cloudy\ and ION to properly calculate those lines in high density, large optical depth gas. 
\item
The predicted (and often observed) short-time changes of the mean emissivity radii of strong emission 
 lines in proportion to \LAGN$^{1/2}$ require corresponding changes in both \Rin\ and \Rout.
\item
The Balmer and Paschen bound-free continua are very strong in all models. Their mean emissivity radii are about half the mean emissivity
radius of the Balmer lines. These continua add substantially to the continuum emission of the central source  
at all wavelengths between 2500 and
10000\AA. This must be taken into account when interpreting the results of continuum RM experiments.
\end{enumerate}

\section{Acknowledgments}

I am grateful to Ari Laor, Alex Baskin, Jonathan Stern, Kirk Korista and Mike Goad for useful comments. I also thank the referee for helping to clarify several
important issues in this work.






\end{document}